%% file: main.tex
\documentclass[12pt]{article}
\usepackage[english]{babel}
\usepackage[utf8]{inputenc}
\usepackage{float}
\usepackage{amsmath,amssymb,graphicx,textcomp}

\newcommand\pubnumber{}
\newcommand\pubdate{\today}

\def\Title#1{\begin{center} {\Large #1 } \end{center}}
\def\Author#1{\begin{center}{ \sc #1} \end{center}}
\def\Address#1{\begin{center}{\footnotesize \it #1} \end{center}}

\newcommand\pubblock{\rightline{\begin{tabular}{l} \pubnumber\\
         \pubdate  \end{tabular}}}
\newenvironment{Abstract}{\begin{quotation}  }{\end{quotation}}
\newenvironment{Presented}{\begin{quotation} \begin{center} 
             PRESENTED AT\end{center}\bigskip 
      \begin{center}\begin{large}}{\end{large}\end{center} \end{quotation}}
\def\Acknowledgements{\bigskip  \bigskip \begin{center} \begin{large}
             \bf ACKNOWLEDGEMENTS \end{large}\end{center}}

\input econfmacros.tex

\newcommand{\RBRC}{
  RIKEN BNL Research Center,
  Brookhaven National Laboratory,
  Upton, New York 11973,
  USA}

\newcommand{\UCONN}{
  Physics Department,
  University of Connecticut,
  Storrs, Connecticut 06269-3046,
  USA}

\newcommand{\NAGOYA}{
  Department of Physics,
  Nagoya University,
  Nagoya 464-8602,
  Japan}
  
\newcommand{\NISHINA}{
  Nishina Center,
  RIKEN,
  Wako, Saitama 351-0198,
  Japan}

\newcommand{\BNL}{
  Physics Department,
  Brookhaven National Laboratory,
  Upton, New York 11973,
  USA}

\newcommand{\CU}{
  Physics Department,
  Columbia University,
  New York, New York 10027,
  USA}

\begin{document}
\begin{titlepage}
\pubblock

\vfill
\Title{Lattice Calculation of the Connected Hadronic Light-by-Light Contribution to the Muon Anomalous Magnetic Moment}
\vfill
\Author{
Luchang Jin\footnote{Speaker},\textsuperscript{3}
Thomas Blum,\textsuperscript{1,2}
Norman Christ,\textsuperscript{3}
Masashi Hayakawa,\textsuperscript{3}
Taku Izubuchi,\textsuperscript{3}
Christoph Lehner,\textsuperscript{3}}

\Address{
\textsuperscript{1}\UCONN\\
\textsuperscript{2}\RBRC\\
\textsuperscript{3}\CU\\
\textsuperscript{4}\NAGOYA\\
\textsuperscript{5}\NISHINA\\
\textsuperscript{6}\BNL\\
}


\vfill
\begin{Abstract}
The anomalous magnetic moment of muon, $g-2$, is a very precisely measured quantity. However, the current measurement disagrees with standard model by about 3 standard deviations. Hadronic vacuum polarization and hadronic light by light are the two types of processes that contribute most to the theoretical uncertainty. I will describe how lattice methods are well-suited to provide a first-principle's result for the hadronic light by light contribution, the various numerical strategies that are presently being used to evaluate it, our current results and the important remaining challenges which must be overcome.
\end{Abstract}
\vfill
\begin{Presented}
Twelfth Conference on the Intersections of Particle and Nuclear Physics\\
Vail Colorado at the Vail Marriott from May 19-24, 2015\\
\end{Presented}
\vfill
\end{titlepage}

\input{body.tex}

\Acknowledgements

We would like to thank our RBC and UKQCD collaborators for helpful discussions and support.
We would also like to thank RBRC for BG/Q computer. This work was supported in part by
US DOE grant DE-SC0011941.

\bibliographystyle{plain}
\bibliography{ref.bib}

\end{document}

%% file: econfmacros.tex
\def\beq{\begin{equation}}
\def\eeq#1{\label{#1}\end{equation}}
\def\eeqn{\end{equation}}

\def\beqa{\begin{eqnarray}}
\def\eeqa#1{\label{#1}\end{eqnarray}}
\def\eeqan{\end{eqnarray}}

\let\bar=\overbar

\def\Dslash{\not{\hbox{\kern-4pt $D$}}}
\def\dslash{\not{\hbox{\kern-2pt $\del$}}}

\def\msb{{\bar{\ssstyle M \kern -1pt S}}}



%% file: body.tex
\section{Introduction}

The anomalous magnetic moment of muon can be defined in terms of the
photon-muon vertex function:
\begin{eqnarray}
  \bar{u} (p') \Gamma_{\nu} (p', p) u (p) & = & \bar{u} (p') \left[ F_1 (q^2)
  \gamma_{\nu} + i \frac{F_2 (q^2)}{4 m_{{\mu}}} [\gamma_{\nu},
  \gamma_{\rho}] q_{\rho} \right] u (p),  \label{vertex-func}
\end{eqnarray}
where $F_2 (0) = (g_{{\mu}} - 2) / 2 \equiv a_{{\mu}}$. The value has
been measured very precisely by BNL E821 {\cite{Bennett:2006fi}}. It can also
be calculated theoretically to great precision as well. {\cite{Blum:2013xva}}
Table \ref{g-2-th-ex} shows various of theoretical contributions to
$a_{{\mu}}$.

\begin{table}[H]
  \begin{center}
    \begin{tabular}{lrr}
      Contribution & $\ensuremath{\operatorname{Value}} \pm
      \ensuremath{\operatorname{Error}}$ & Ref\\
      &  & \\
      QED incl. 5-loops & $116584718.951 \pm 0.080$ & {\cite{Aoyama:2012wk}}\\
      HVP LO & $6923 \pm 42$ & {\cite{Davier:2010nc}}\\
      & $6949 \pm 43$ & {\cite{Hagiwara:2011af}}\\
      HVP NLO & $- 98.4 \pm 0.7$ & {\cite{Hagiwara:2011af}}\\
      Hadronic Light by Light & $105 \pm 26$ & {\cite{Prades:2009tw}}\\
      Weak incl. 2-loops & $153.6 \pm 1.0$ & {\cite{Gnendiger:2013pva}}\\
      Standard Model & $116591802 \pm 49$ & {\cite{Davier:2010nc}}\\
      & $116591828 \pm 50$ & {\cite{Hagiwara:2011af}}\\
      Experiment (0.54 ppm) & $116592089 \pm 63$ & {\cite{Bennett:2006fi}}\\
      &  & \\
      Difference
      ($\ensuremath{\operatorname{Exp}}-\ensuremath{\operatorname{SM}}$) &
      $287 \pm 80$ & {\cite{Davier:2010nc}}\\
      & $261 \pm 78$ & {\cite{Hagiwara:2011af}}
    \end{tabular}
  \end{center}
  
  \
  \caption{\label{g-2-th-ex}Comparison between standard model theory and
  experiment. [in units of $10^{- 11}$]}
\end{table}

\begin{figure}[H]
  \begin{center}
    \resizebox{0.3\columnwidth}{!}{\includegraphics{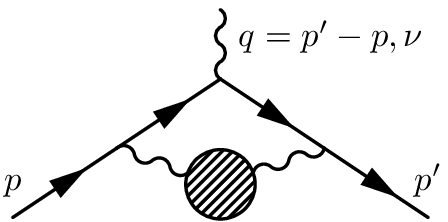}}
    \resizebox{0.3\columnwidth}{!}{\includegraphics{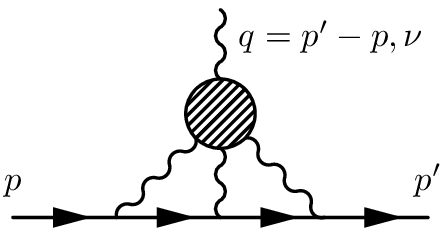}}
  \end{center}
  \caption{\label{hvp-hlbl}(Left) Hadronic vacuum polarization diagram.
  (Right) Hadronic light-by-light diagram.}
\end{figure}

The around three standard deviations between the experiment and theory makes
muon $g - 2$ a very interesting quantity. A much more accurate experiment by
Fermilab E989 is expected in a few years, so a more accurate theoretical
determination would be necessary. Figure \ref{hvp-hlbl} shows the two diagrams
that are the major sources of the theoretical uncertainty.

In this paper, we will only discuss the lattice calculation of connected
hadronic light-by-light amplitude. This subject was begun by T. Blum, M.
Hayakawa, and T. Izubuchi more than 5 years ago
{\cite{Hayakawa:2005eq,Blum:2014oka}}. We have improved the
methodology dramatically recently with three major changes. First, we
calculate the process at $\mathcal{O} (\alpha^3)$ with six explicitly internal
QED interaction vertices, so no lower order noises or higher order systematic
errors will affect our results. Second, we do not generate stochastic QED
gauge field configurations, all the photon propagators are calculated exactly
based on analytic expressions and Fourier transformations. Third, we compute
$F_2 (q^2 = 0)$ directly in finite volume. A much more accurate result is
obtained with the improved method. We then applied this method with
simulations parameters closer to physical kinematics.

\section{Evaluation Strategy}

We start the discussion by spell out the complete expression of the connected
light-by-light diagram.

\begin{figure}[H]
  \begin{center}
    \resizebox{0.35\columnwidth}{!}{\includegraphics{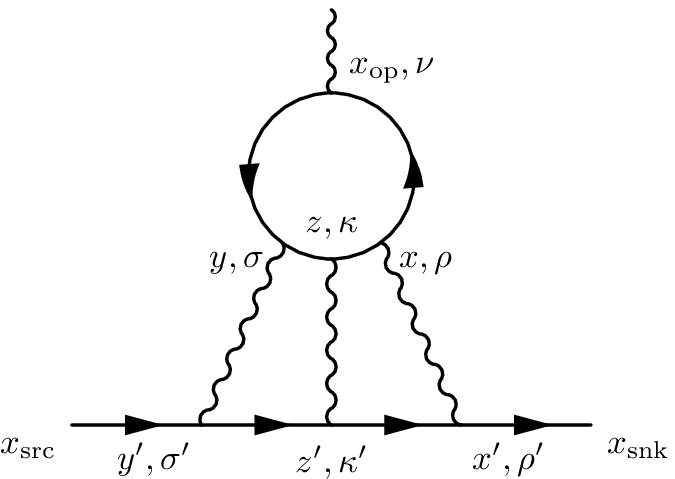}}
    \resizebox{0.35\columnwidth}{!}{\includegraphics{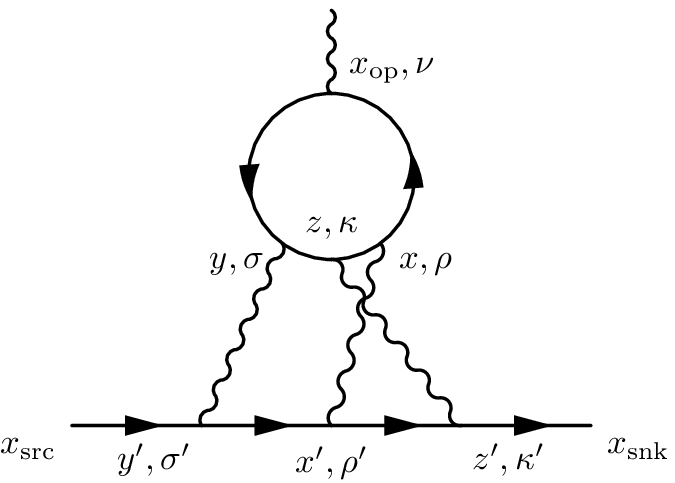}}
  \end{center}
  \caption{\label{lbl}Light-by-Light diagrams. There are 4 other possible
  permutations.}
\end{figure}

We denote the momentum carried by the external photon by $\mathbf{q}$. Also,
we use Breit-frame, so the initial and final muon states have exactly the same
energy,
\begin{eqnarray}
  E_{\mathbf{q}/ 2} & = & \sqrt{(\mathbf{q}/ 2)^2 + m_{{\mu}}^2} . 
\end{eqnarray}
To project each of initial and final states onto a single particle state, we
need to take the limits $t_{\text{src}} \rightarrow - \infty$ and
$t_{\text{snk}} \rightarrow \infty$. Under these limits, the amplitude in
momentum space can be described by the form factors
\begin{eqnarray}
  \mathcal{M}^{\text{LbL}}_{\nu} (\mathbf{q}) & = & e^{i\mathbf{q} \cdot
  \mathbf{x}_{\text{op}}}  \sum_{\mathbf{x}_{\text{snk}},
  \mathbf{x}_{\text{src}}} e^{- i\mathbf{q}/ 2 \cdot \left(
  \mathbf{x}_{\text{src}} +\mathbf{x}_{\text{snk}} \right)} e^{E_{\mathbf{q}/
  2} t_{\text{sep}}} \nonumber\\
  & \cdot & S_{{\mu}} \left( x_{\text{snk}}, x_{\text{op}} \right) 
  \left[ F_1 (q^2) \gamma_{\nu} + i \frac{F_2 (q^2)}{4 m} [\gamma_{\nu},
  \gamma_{\rho}] q_{\rho} \right] S_{{\mu}} \left( x_{\text{op}},
  x_{\text{src}} \right),  \label{lbl-ffs}
\end{eqnarray}
where $t_{\text{sep}} = t_{\text{snk}} - t_{\text{src}}$. The above expression
is independent of $x_{\ensuremath{\operatorname{op}}}$, and
$\mathcal{M}^{\text{LbL}}$ is only a function of $\mathbf{q}$ as one would
expect. In terms of Feynman diagrams, the amplitude is
\begin{eqnarray}
  \mathcal{M}^{\text{LbL}}_{\nu} (\mathbf{q}) & = & e^{i\mathbf{q} \cdot
  \mathbf{x}_{\text{op}}}  \sum_{\mathbf{x}_{\text{snk}},
  \mathbf{x}_{\text{src}}} e^{- i\mathbf{q}/ 2 \cdot \left(
  \mathbf{x}_{\text{src}} +\mathbf{x}_{\text{snk}} \right)} e^{E_{\mathbf{q}/
  2} t_{\text{sep}}} \mathcal{M}^{\text{LbL}}_{\nu} \left( x_{\text{op}},
  x_{\text{snk}}, x_{\text{src}} \right), 
\end{eqnarray}
\begin{eqnarray}
  \mathcal{M}^{\text{LbL}}_{\nu} \left( x_{\text{op}}, x_{\text{snk}},
  x_{\text{src}} \right) & = & \sum_{x, y, z} \mathcal{F}_{\nu} \left( x, y,
  z, x_{\text{op}}, x_{\text{snk}}, x_{\text{src}} \right), 
\end{eqnarray}
\begin{eqnarray}
  &  & \mathcal{F}_{\nu} \left( x, y, z, x_{\text{op}}, x_{\text{snk}},
  x_{\text{src}} \right) \nonumber\\
  & = & - (- i e)^6 \sum_{q = u, d, s} (e_q / e)^4  \left\langle
  \ensuremath{\operatorname{tr}} \left[ \gamma_{\rho} S_q (x, z)
  \gamma_{\kappa} S_q (z, y) \gamma_{\sigma} S_q \left( y, x_{\text{op}}
  \right) \gamma_{\nu} S_q \left( x_{\text{op}}, x \right) \right]
  \right\rangle_{\text{QCD}} \nonumber\\
  & \cdot & \sum_{x', y', z'} G_{\rho \rho'} (x, x') G_{\sigma \sigma'} (y,
  y') G_{\kappa \kappa'} (z, z') \nonumber\\
  & \cdot & \left[ S_{{\mu}} \left( x_{\text{snk}}, x' \right)
  \gamma_{\rho'} S_{{\mu}} (x', z') \gamma_{\kappa'} S_{{\mu}} (z',
  y') \gamma_{\sigma'} S_{{\mu}} \left( y', x_{\text{src}} \right) \right.
  \nonumber\\
  &  & + S_{{\mu}} \left( x_{\text{snk}}, z' \right) \gamma_{\kappa'}
  S_{{\mu}} (z', x') \gamma_{\rho'} S_{{\mu}} (x', y')
  \gamma_{\sigma'} S_{{\mu}} \left( y', x_{\text{src}} \right) \nonumber\\
  &  & \left. + \text{other 4 permutations} \right],  \label{lbl-amp}
\end{eqnarray}
where $e_u / e = 2 / 3$, $e_d / e = e_s / e = - 1 / 3$.

It is very difficult to evaluate the above complicated three-loop formula
directly on the lattice, because we can not afford the $\mathcal{O} \left(
\text{Volume}^2 \right)$ complexity. We need some stochastic method to
evaluate the above formula. In Ref
{\cite{Hayakawa:2005eq,Blum:2014oka}} we evaluated the quark and muon
propagators in the background of quenched QED fields. This will generate all
kinds of diagrams, a nice subtraction scheme is then used to subtract all the
unwanted pieces, except some higher order terms suppressed by additional
powers of $\alpha$.

\begin{figure}[H]
  \begin{center}
    \begin{tabular}{cccc}
      \resizebox{!}{0.17\columnwidth}{\includegraphics{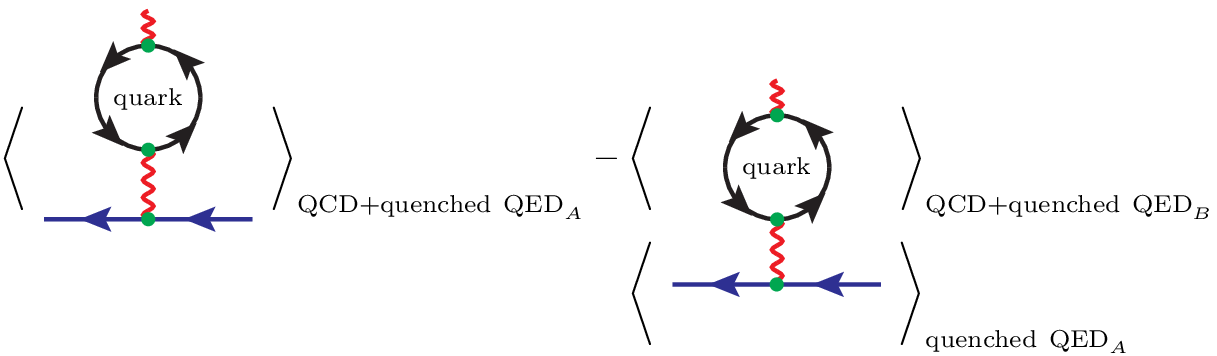}}
      & = & $3 \times$ &
      \resizebox{!}{0.17\columnwidth}{\includegraphics{fig-35.eps}}
    \end{tabular}
  \end{center}
  \caption{One typical diagram remains after subtraction is shown on the left,
  5 others are not shown. See Ref
  {\cite{Hayakawa:2005eq,Blum:2014oka}} for details.}
\end{figure}

Although the central value of the lower order terms is subtracted completely,
the noise terms is not. After subtraction, the noise is on the order of
$\mathcal{O} (e^4)$ compare with the signal, which is on the order of
$\mathcal{O} (e^6)$. This lower order noise problem can be solved by inserting
the stochastic photon explicitly using the sequential source method
{\cite{Jin:2014cea}}. Then we would be only evaluating the connected HLbL
diagram, without higher order error or lower order noise.

\begin{figure}[H]
  \begin{center}
    \resizebox{0.35\columnwidth}{!}{\includegraphics{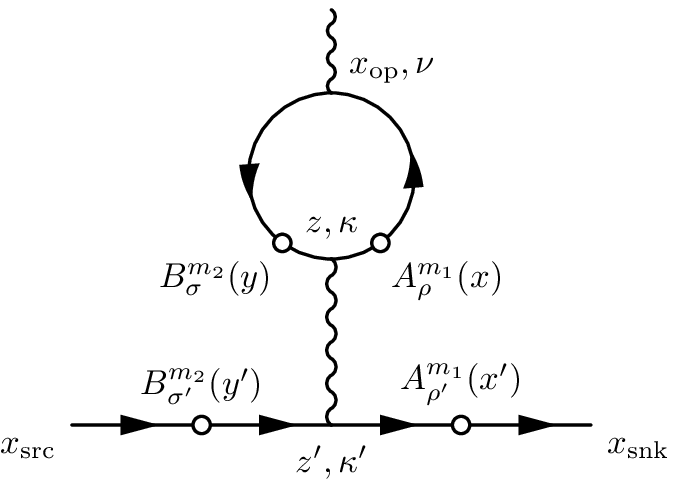}}
    \resizebox{0.35\columnwidth}{!}{\includegraphics{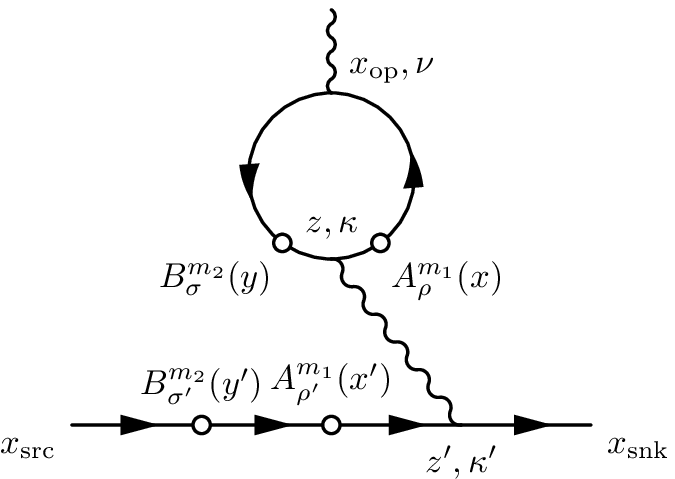}}
  \end{center}
  \caption{Light-by-Light diagrams calculated with one exact photon and two
  stochastic photons. There are 4 other possible permutations. See Ref
  {\cite{Jin:2014cea}} for details.}
\end{figure}

\subsection{Point Source Photon Method}

However, there is still a serious problem in the approach above, that is the
noise will increase in larger volume. We will call it ``disconnected-diagram''
problem, because it is similar to the problem one usually encounter when
computing diagrams with two or more parts that are not connected by fermion
lines on lattice. In our case, the quark loop and the muon line are
``disconnected'', so in large volume, lots of noise will be generated from the
region where $x'$ is far from $x$ or $y'$ is far from $y$, since this noise is
not suppressed at all.

Use of an analytical photon propagator it contains instead of a stochastic
photon field would solve this ``disconnected-diagram'' problem, but it is not
possible to exactly evaluation the full expression because it contains too
many loops. As a trade-off, we use two point source photons with sources at
$x$ and $y$, which will be chosen randomly. The $8 L^4$-dimensional stochastic
integral over E\&M fields will be replaced by a very standard 8-dimensional
Monte Carlo integral over two space-time points, and the integrand only
depends on the relative position of the two points after QCD ensemble average.
To achieve this, we rearrange the expression of the amplitude by defining
\begin{eqnarray}
  \mathcal{F}_{\nu} \left( \mathbf{q}, x, y, z, x_{\text{op}} \right) & = &
  \sum_{\mathbf{x}_{\text{snk}}, \mathbf{x}_{\text{src}}} e^{- i\mathbf{q}/ 2
  \cdot \left( \mathbf{x}_{\text{src}} +\mathbf{x}_{\text{snk}} \right)}
  e^{E_{\mathbf{q}/ 2} t_{\text{sep}}} \mathcal{F}_{\nu} \left( x, y, z,
  x_{\text{op}}, x_{\text{snk}}, x_{\text{src}} \right) . 
\end{eqnarray}
Then, we will have
\begin{eqnarray}
  \mathcal{M}^{\text{LbL}}_{\nu} (\mathbf{q}) & = & e^{i\mathbf{q} \cdot
  \mathbf{x}_{\text{op}}}  \sum_{x, y, z} \mathcal{F}_{\nu} \left( \mathbf{q},
  x, y, z, x_{\text{op}} \right) . 
\end{eqnarray}
Translational invariance of $\mathcal{F}_{\nu} \left( x, y, z, x_{\text{op}},
x_{\text{snk}}, x_{\text{src}} \right)$ leads to the following equation
\begin{eqnarray}
  &  & e^{i\mathbf{q} \cdot \mathbf{x}_{\text{op}}} \mathcal{F}_{\nu} \left(
  \mathbf{q}, x, y, z, x_{\text{op}} \right) \nonumber\\
  & = & e^{i\mathbf{q} \cdot \left( \mathbf{x}_{\text{op}} -
  \frac{\mathbf{x}+\mathbf{y}}{2} \right)} \mathcal{F}_{\nu} \left(
  \mathbf{q}, \frac{x - y}{2}, - \frac{x - y}{2}, z - \frac{x + y}{2},
  x_{\text{op}} - \frac{x + y}{2} \right) . 
\end{eqnarray}
Therefore
\begin{eqnarray}
  \mathcal{M}^{\text{LbL}}_{\nu} (\mathbf{q}) & = & \sum_{x, y, z}
  e^{i\mathbf{q} \cdot \left( \mathbf{x}_{\text{op}} -
  \frac{\mathbf{x}+\mathbf{y}}{2} \right)} \mathcal{F}_{\nu} \left(
  \mathbf{q}, \frac{x - y}{2}, - \frac{x - y}{2}, z - \frac{x + y}{2},
  x_{\text{op}} - \frac{x + y}{2} \right) \nonumber\\
  & = & \sum_r \left[ \sum_{\tilde{z}, \tilde{x}_{\text{op}}} e^{i\mathbf{q}
  \cdot \tilde{\mathbf{x}}_{\text{op}}} \mathcal{F}_{\nu} \left( \mathbf{q},
  \frac{r}{2}, - \frac{r}{2}, \tilde{z}, \tilde{x}_{\text{op}} \right) \right]
  . 
\end{eqnarray}
where $r = x - y$, $\tilde{z} = z - (x + y) / 2$, and $\tilde{x}_{\text{op}} =
x_{\text{op}} - (x + y) / 2$. This would be the formula suitable for our
proposed strategy. The inner sums over $\tilde{z}$ and $\tilde{x}_{\text{op}}$
can be easily summed over the entire lattice as sinks with point source
propagators originated at $x$ and $y$. The outer sum will be performed by
random sampling the $x$ and $y$ positions. It should be noted that the
integrand is sharply peaked in the small $r$ region, so one should sample this
region more frequently. In fact, we choose to compute all possible\footnote{Up
to discrete symmetries, e.g. reflections.} $r$ less than certain limit
$r_{\text{max}}$ and simply add them together as the ``short distance''
contribution. Then, we only randomly sample the region where $r >
r_{\text{max}}$, with some probability distribution tailored for the specific
pion mass.

There is also a possible $M^2$ trick, similar to the one described in Ref
{\cite{Jin:2014cea}}, which can be applied to this setup in a similar fashion.
This trick works as follows. First, one chooses a random point as the
reference point $x_{\text{ref}}$. Second, one chooses a set of $M$ points
$c_1$, $c_2$ ... $c_M$ around $x_{\ensuremath{\operatorname{ref}}}$ with some
pre-specified probability distribution $q (| c_i -
x_{\ensuremath{\operatorname{ref}}} |)$. Then, the distance between any two
points $x$ and $y$ within this set is given by
\begin{eqnarray}
  p (| x - y |) & = & \sum_{x_{\text{ref}}} q \left( \left| x - x_{\text{ref}}
  \right| \right) q \left( \left| y - x_{\text{ref}} \right| \right) . 
\end{eqnarray}
With this trick, we obtained $M (M - 1) / 2$ point pairs by just computing $M$
point source propagators. We have experimented this method on our $32^3$ $4.6
\mathrm{\ensuremath{\operatorname{fm}}}$ lattice {\cite{Arthur:2012yc}} with a
$171 \mathrm{\ensuremath{\operatorname{MeV}}}$ pion and a $134
\mathrm{\ensuremath{\operatorname{MeV}}}$ muon using $M = 16$. Under this
setting, we found that this $M^2$ trick is very effective, all the pairs are
almost statistically independent even though they are just different
combinations out of the same set of points on the same configuration near the
same reference point $x_{\text{ref}}$. This trick can also be applied after
including the two following improvements. However, we didn't use it in our
recent numerical studies, because the light quark inversion is made very fast
by using Mobius/zMobius fermions {\cite{Brower:2012vk}}, the AMA
{\cite{Blum:2012uh}} technique, and efficient code {\cite{Boyle:2009vp}}.
Comparatively, the muon part of the computation, which would need to be
performed $M (M - 1) / 2$ times should we use this $M^2$ trick, is quite
expensive.

\subsection{Conserved External Current Improvement}

\begin{figure}[H]
  \begin{center}
    \resizebox{0.25\columnwidth}{!}{\includegraphics{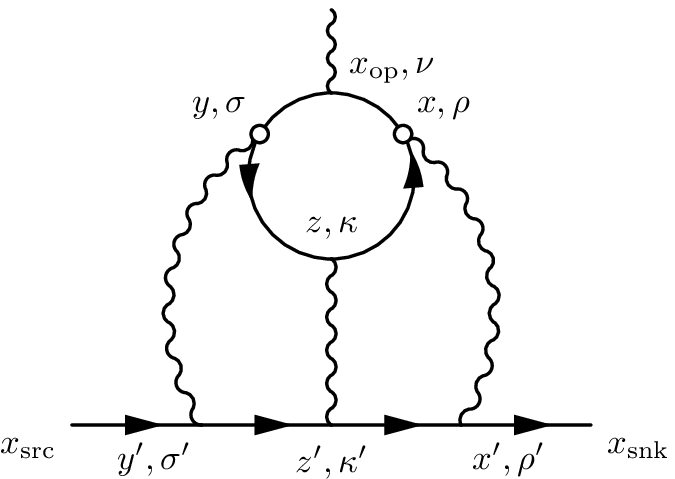}}
    \resizebox{0.25\columnwidth}{!}{\includegraphics{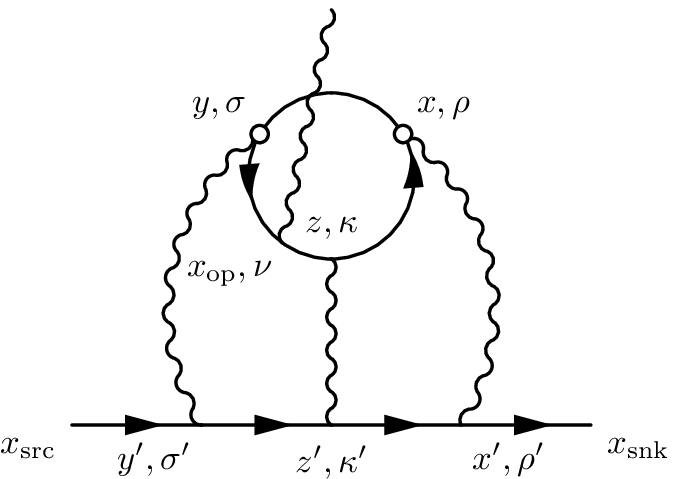}}
    \resizebox{0.25\columnwidth}{!}{\includegraphics{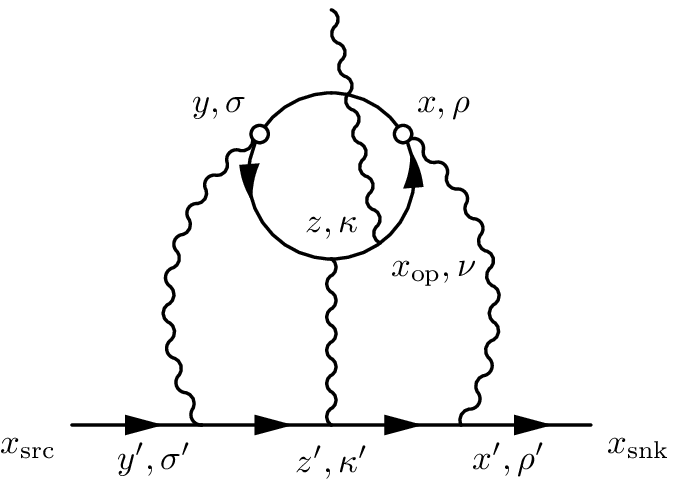}}
  \end{center}
  \caption{\label{fig-c-3}Diagrams showing the three different possible
  insertions of the external photon when the vertices $x$ and $y$ are fixed.
  For each of these three diagrams there are five other possible permutations
  of the connections between the three internal photons and the muon line that
  are not shown. The contributions of each of these three sets of six
  contractions will be the same after the stochastic average over the vertices
  $x$ and $y$. (Left) This is the diagram that we have already calculated.
  (Middle) We need to compute sequential source propagators at $x_{\text{op}}$
  for each polarizations of the external photon. (Right) We also need to
  compute sequential source propagators at $x_{\text{op}}$, but with the
  external photon momentum in opposite direction, since we need use
  $\gamma_5$-hermiticity to reverse the direction of the propagators, which
  reverses the momentum of the external photon as well.}
\end{figure}

With the point source photon method, the noise for
$\mathcal{M}^{\text{LbL}}_{\nu} (\mathbf{q})$ should stay relative constant
when we increase the volume of the lattice. However, it should be noticed that
the signal of the anomalous magnetic moment is proportion to $q$, so the
signal to noise ratio problem in large volume limit remains unless we can
control the noise to be proportion to $q$ as well. Recall the reason that the
signal proportion to $q$ is the Ward identity, so we should enforce the Ward
identity configuration by configuration by including contributions from all
the possible external photon insertions, and use the lattice conserved current
at $x_{\text{op}}$. After including the three types of diagrams shown in
Figure \ref{fig-c-3}, we have
\begin{eqnarray}
  &  & \mathcal{F}^C_{\nu} \left( x, y, z, x_{\text{op}}, x_{\text{snk}},
  x_{\text{src}} \right) \nonumber\\
  &  & \hspace{1cm} = \frac{1}{3} \mathcal{F}_{\nu} \left( x, y, z,
  x_{\text{op}}, x_{\text{snk}}, x_{\text{src}} \right) + \frac{1}{3}
  \mathcal{F}_{\nu} \left( y, z, x, x_{\text{op}}, x_{\text{snk}},
  x_{\text{src}} \right) \nonumber\\
  &  & \hspace{2cm} + \frac{1}{3} \mathcal{F}_{\nu} \left( z, x, y,
  x_{\text{op}}, x_{\text{snk}}, x_{\text{src}} \right),  \label{lbl-amp-c}
\end{eqnarray}
\begin{eqnarray}
  \mathcal{F}^C_{\nu} \left( \mathbf{q}, x, y, z, x_{\text{op}} \right) & = &
  \sum_{\mathbf{x}_{\text{snk}}, \mathbf{x}_{\text{src}}} e^{- i\mathbf{q}/ 2
  \cdot \left( \mathbf{x}_{\text{src}} +\mathbf{x}_{\text{snk}} \right)}
  e^{E_{\mathbf{q}/ 2} t_{\text{sep}}} \mathcal{F}^C_{\nu} \left( x, y, z,
  x_{\text{op}}, x_{\text{snk}}, x_{\text{src}} \right) . 
\end{eqnarray}
We then have a similar formula
\begin{eqnarray}
  \mathcal{M}^{\text{LbL}}_{\nu} (\mathbf{q}) & = & e^{i\mathbf{q} \cdot
  \mathbf{x}_{\text{op}}}  \sum_{x, y, z} \mathcal{F}^C_{\nu} \left(
  \mathbf{q}, x, y, z, x_{\text{op}} \right) \nonumber\\
  & = & \sum_r \left[ \sum_{\tilde{z}, \tilde{x}_{\text{op}}} e^{i\mathbf{q}
  \cdot \tilde{\mathbf{x}}_{\text{op}}} \mathcal{F}^C_{\nu} \left( \mathbf{q},
  \frac{r}{2}, - \frac{r}{2}, \tilde{z}, \tilde{x}_{\text{op}} \right) \right]
  .  \label{lbl-m-f-c}
\end{eqnarray}
There is a side effect of including all three possible external photon
insertions. According to the definition Eq. (\ref{lbl-amp-c}), we have
$\mathcal{F}^C_{\nu} \left( \mathbf{q}, x, y, z, x_{\text{op}} \right)
=\mathcal{F}^C_{\nu} \left( \mathbf{q}, y, z, x, x_{\text{op}} \right)
=\mathcal{F}^C_{\nu} \left( \mathbf{q}, z, x, y, x_{\text{op}} \right)$. This
allows us to apply another trick,
\begin{eqnarray}
  \sum_{x, y, z} \mathcal{F}^C_{\nu} \left( \mathbf{q}, x, y, z, x_{\text{op}}
  \right) & = & \sum_{x, y, z} \mathfrak{Z}\mathcal{F}^C_{\nu} \left(
  \mathbf{q}, x, y, z, x_{\text{op}} \right), 
\end{eqnarray}
where
\begin{eqnarray}
  \mathfrak{Z} & = & \left\{ \begin{array}{ll}
    3 & \text{if } | x - y | < | x - z | \text{ and } | x - y | < | y - z |\\
    3 / 2 & \text{if } | x - y | = | x - z | < | y - z | \text{ or } | x - y |
    = | y - z | < | x - z |\\
    1 & \text{if } | x - y | = | x - z | = | y - z |\\
    0 & \text{otherwise}
  \end{array} \right. . 
\end{eqnarray}
Following Eq. (\ref{lbl-m-f-c}), we obtain
\begin{eqnarray}
  \mathcal{M}^{\text{LbL}}_{\nu} (\mathbf{q}) & = & \sum_r \left[
  \sum_{\tilde{z}} \mathfrak{Z} \sum_{\tilde{x}_{\text{op}}} e^{i\mathbf{q}
  \cdot \tilde{\mathbf{x}}_{\text{op}}} \mathcal{F}^C_{\nu} \left( \mathbf{q},
  \frac{r}{2}, - \frac{r}{2}, \tilde{z}, \tilde{x}_{\text{op}} \right) \right]
  . 
\end{eqnarray}
This trick further suppresses contributions from large $| r |$, where most
noise would enter, and makes the importance sampling and complete summation of
the short distance region more effective.

\subsection{Zero External Momentum Transfer Improvement}

Recall the basic form of the Ward identity
\begin{eqnarray}
  \Delta^{\ast}_{\left( x_{\text{op}} \right)_{\nu}} S_q \left( b,
  x_{\text{op}} \right) \gamma_{\nu} S_q \left( x_{\text{op}}, a \right) & = &
  S_q (b, a) \left[ \delta \left( x_{\text{op}} - a \right) - \delta \left(
  x_{\text{op}} - b \right) \right] . 
\end{eqnarray}
Note that the $\gamma_{\nu}$ should be interpreted as a lattice version
conserved current at $x_{\text{op}}$. The current conservation formulae
\begin{eqnarray}
  \Delta^{\ast}_{\left( x_{\text{op}} \right)_{\nu}} F^C_{\nu} \left( x, y, z,
  x_{\text{op}}, x_{\text{snk}}, x_{\text{src}} \right) & = & 0, 
\end{eqnarray}
\begin{eqnarray}
  \Delta^{\ast}_{\left( x_{\text{op}} \right)_{\nu}}  \mathcal{F}^C_{\nu}
  \left( \mathbf{q}, x, y, z, x_{\text{op}} \right) & = & 0, 
\end{eqnarray}
will be true configuration by configuration without discretization error or
finite volume error, provided we include all possible external photon
insertions and use the lattice version of the conserved current at
$x_{\text{op}}$.

With the above formula, we can prove $\sum_{\tilde{x}_{\text{op}}} F_{\nu}^C$
vanishes except for terms suppressed exponentially by the lattice size. The
reason is that the net total sum of a localized conserved current has to
vanish. Mathematically, $F^C_{\nu}$ is exponentially suppressed at large
$\tilde{x}_{\text{op}}$, therefore we can ignore the surface term in a
sufficiently large volume,
\begin{eqnarray}
  0 & = & \sum_{\tilde{x}_{\text{op}}} \Delta^{\ast}_{\left(
  \tilde{x}_{\text{op}} \right)_{\rho}}  \left[ \left( \tilde{x}_{\text{op}}
  \right)_{\nu} \mathcal{F}^C_{\rho} \left( \mathbf{q}, \frac{r}{2}, -
  \frac{r}{2}, \tilde{z}, \tilde{x}_{\text{op}} \right) \right] \nonumber\\
  & = & \sum_{\tilde{x}_{\text{op}}}  \left[ \mathcal{F}^C_{\nu} \left(
  \mathbf{q}, \frac{r}{2}, - \frac{r}{2}, \tilde{z}, \tilde{x}_{\text{op}}
  \right) + \left( \tilde{x}_{\text{op}} \right)_{\nu} \Delta^{\ast}_{\left(
  \tilde{x}_{\text{op}} \right)_{\rho}} \mathcal{F}^C_{\rho} \left(
  \mathbf{q}, \frac{r}{2}, - \frac{r}{2}, \tilde{z}, \tilde{x}_{\text{op}}
  \right) \right] \nonumber\\
  & = & \sum_{\tilde{x}_{\text{op}}} \mathcal{F}^C_{\nu} \left( \mathbf{q},
  \frac{r}{2}, - \frac{r}{2}, \tilde{z}, \tilde{x}_{\text{op}} \right), 
\end{eqnarray}
In infinite volume, we can safely subtract above term from our amplitude, we
obtain
\begin{eqnarray}
  \mathcal{M}^{\text{LbL}}_{\nu} (\mathbf{q}) & = & \sum_r \left[
  \sum_{\tilde{z}, \tilde{x}_{\text{op}}} \left( e^{i\mathbf{q} \cdot
  \tilde{\mathbf{x}}_{\text{op}}} - 1 \right) \mathcal{F}^C_{\nu} \left(
  \mathbf{q}, \frac{r}{2}, - \frac{r}{2}, \tilde{z}, \tilde{x}_{\text{op}}
  \right) \right] . 
\end{eqnarray}
Now, the above expression vanishes explicitly when $q \rightarrow 0$. The
leading order contribution is
\begin{eqnarray}
  \mathcal{M}^{\text{LbL}}_{\nu} (\mathbf{q}) & = & \sum_r \left[
  \sum_{\tilde{z}, \tilde{x}_{\text{op}}} i\mathbf{q} \cdot
  \tilde{\mathbf{x}}_{\text{op}} \mathcal{F}^C_{\nu} \left(
  \ensuremath{\boldsymbol{0}}, \frac{r}{2}, - \frac{r}{2}, \tilde{z},
  \tilde{x}_{\text{op}} \right) \right] +\mathcal{O} (q^2) . 
\end{eqnarray}
Matching with Eq. (\ref{lbl-ffs}), we can see that
\begin{eqnarray}
  F_1 (0) & = & 0, 
\end{eqnarray}
\begin{eqnarray}
  &  & \sum_{\mathbf{x}_{\text{snk}}, \mathbf{x}_{\text{src}}}
  e^{m_{{\mu}} t_{\text{sep}}} S_{{\mu}} \left( x_{\text{snk}},
  x_{\text{op}} \right)  \left[ i \frac{F_2 (0)}{4 m} [\gamma_{\nu},
  \gamma_{\rho}] q_{\rho} \right] S_{{\mu}} \left( x_{\text{op}},
  x_{\text{src}} \right)  \label{f2-x-f-c}\\
  &  & \hspace{5cm} = \sum_r \left[ \sum_{\tilde{z}, \tilde{x}_{\text{op}}}
  i\mathbf{q} \cdot \tilde{\mathbf{x}}_{\text{op}} \mathcal{F}^C_{\nu} \left(
  \ensuremath{\boldsymbol{0}}, \frac{r}{2}, - \frac{r}{2}, \tilde{z},
  \tilde{x}_{\text{op}} \right) \right] . \nonumber
\end{eqnarray}
Now we have an explicit formula about $F_2$ at zero momentum transfer.
Although the expression is derived in infinite volume, we can still evaluate
it in finite volume on the lattice only subject to normal power law
finite-volume effects just like other lattice computations include QED.
Similarly, although the derivation of Eq. (\ref{f2-x-f-c}) assumes strict
current conservation at $x_{\text{op}}$ guaranteed by the lattice version of
the conserved current, the final form of the Eq. (\ref{f2-x-f-c}) has no
superficial divergence in the ultra-violet region, thus one can also use a
local current at $x_{\text{op}}$. Finally, we further simplify the above
expression by cancelling $q$ on both sides of the equation,
\begin{eqnarray}
  &  & \frac{F_2 (0)}{m}  \sum_{\mathbf{x}_{\text{snk}},
  \mathbf{x}_{\text{src}}} e^{m_{{\mu}} t_{\text{sep}}} S_{{\mu}}
  \left( x_{\text{snk}}, x_{\text{op}} \right)  \frac{\vec{\Sigma}}{2}
  S_{{\mu}} \left( x_{\text{op}}, x_{\text{src}} \right) \nonumber\\
  &  & \hspace{2cm} = \sum_r \left[ \sum_{\tilde{z}, \tilde{x}_{\text{op}}}
  \frac{1}{2}  \tilde{\mathbf{x}}_{\text{op}} \times i \vec{\mathcal{F}}^C
  \left( \ensuremath{\boldsymbol{0}}, \frac{r}{2}, - \frac{r}{2}, \tilde{z},
  \tilde{x}_{\text{op}} \right) \right] \nonumber\\
  &  & \hspace{2cm} = \sum_r \left[ \sum_{\tilde{z}} \mathfrak{Z}
  \sum_{\tilde{x}_{\text{op}}} \frac{1}{2}  \tilde{\mathbf{x}}_{\text{op}}
  \times i \vec{\mathcal{F}}^C \left( \ensuremath{\boldsymbol{0}},
  \frac{r}{2}, - \frac{r}{2}, \tilde{z}, \tilde{x}_{\text{op}} \right)
  \right], 
\end{eqnarray}
where $\Sigma_i = \frac{1}{4 i} \epsilon_{i j k} [\gamma_j, \gamma_k]$. The
evaluation strategy is the same as before. The sum over
$\tilde{x}_{\text{op}}$ is performed by the sequential source
method.\footnote{For each of the two points in the point pair, we need to
compute $1$ point source propagator and $3$ sequential source propagators for
$3$ magnetic moment directions. If $M^2$ trick is applied, for each point we
need to compute $3$ additional sequential source propagators to shift the
origin of $x_{\ensuremath{\operatorname{op}}}$, then we can combine the points
in arbitrary ways.} The sum over $\tilde{z}$ is easily evaluated because
$\tilde{z}$ is a sink. Again, the final sum over $r$ is performed by random
sampling $r$ according to a probability distribution $p (r)$, except for $r
\leqslant r_{\text{max}}$ in which case we compute all possible $r$ up to
discrete symmetries and sum them with appropriate multiplicity factors.

\section{Numerical Studies}

In this section we describe three studies. We start by presenting the QED
test, computing muon leptonic light-by-light process on lattice and also study
the finite volume effects. We then present our $24^3$ lattice simulations,
which we compare with our previous study in Ref {\cite{Blum:2014oka}}.
Finally, we apply our new evaluation strategy to a $32^3$ lattice.\footnote{At
the conference, we presentd results from the $24^3$ simulation at zero
momentum transfer using the moment method in Eq. (\ref{f2-x-f-c}). The
zero-momentum transfer results for the muon leptonic simulation and the
$32^3$, $m_{\pi} = 171 \mathrm{\ensuremath{\operatorname{MeV}}}$ simulation
were obtained later. At the conference we reported results with non-zero
momentum transfer for these two studies.}

\subsection{Muon Leptonic Light-by-Light and Finite Volume Effects}

We start by using the method described above to compute the muon leptonic
light-by-light process. The computation is performed on three different
physical volumes and each with three different lattice spacings. The lattice
spacing is determined by the physical muon mass, $m_{{\mu}} = 106
\mathrm{\ensuremath{\operatorname{MeV}}}$.

\begin{figure}[H]
  \begin{center}
    \resizebox{0.4\columnwidth}{!}{\includegraphics{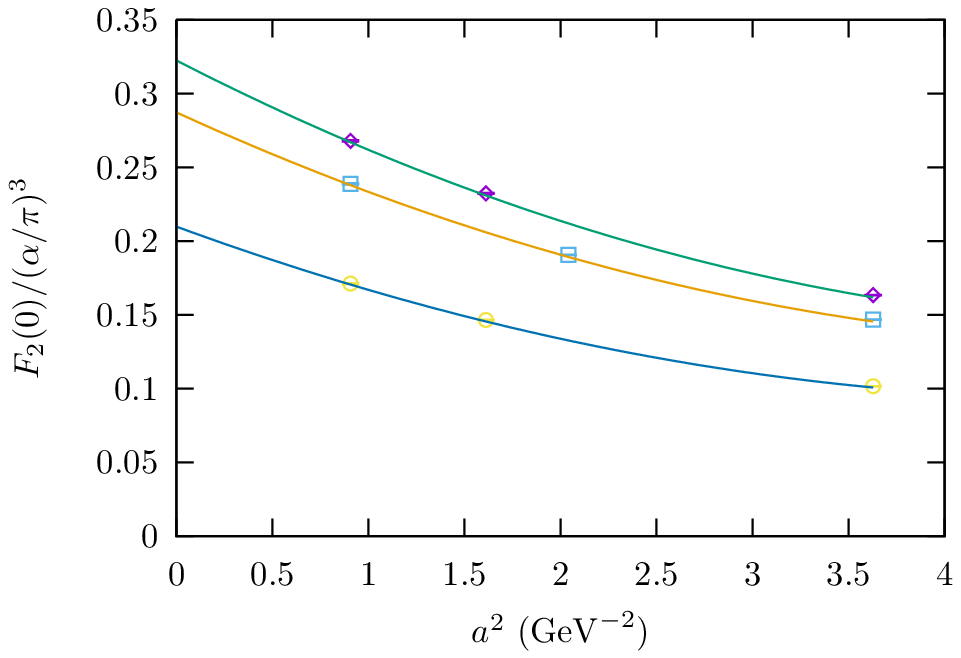}}
    \resizebox{0.4\columnwidth}{!}{\includegraphics{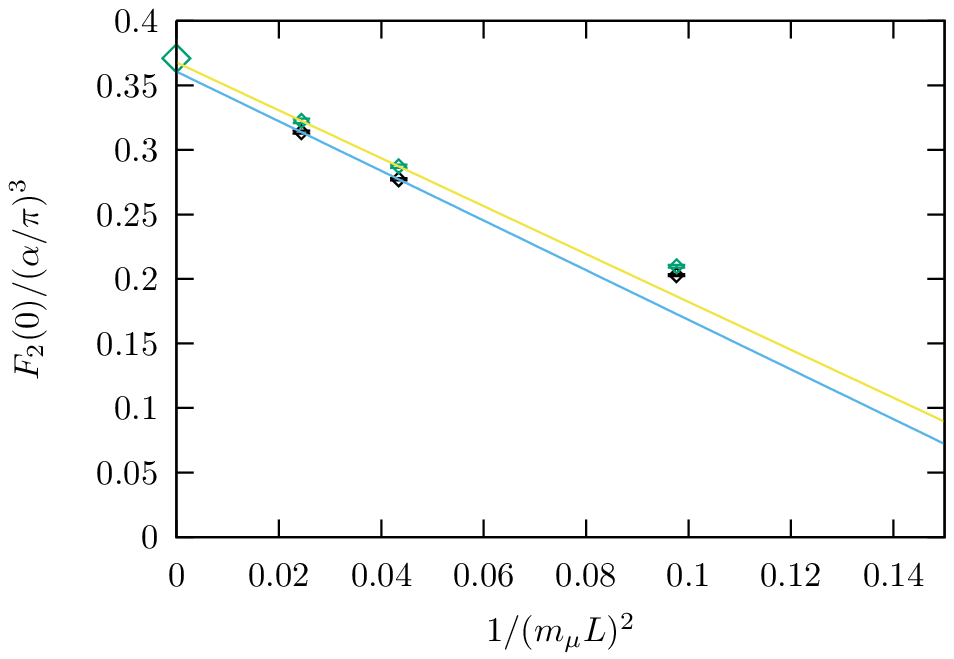}}
  \end{center}
  \caption{Muon leptonic light by light contribution to muon $g - 2$. (Left)
  Plots of the continuum extrapolation for three different physical lattice
  sizes $L = 11.9 \mathrm{\ensuremath{\operatorname{fm}}}$ (diamonds), $8.9
  \mathrm{\ensuremath{\operatorname{fm}}}$ (squares) and $5.9
  \mathrm{\ensuremath{\operatorname{fm}}}$ (circles) by assigning
  $m_{{\mu}} = 106 \mathrm{\ensuremath{\operatorname{MeV}}}$. The curves
  shown are quadratic functions of $a^2$ chosen to pass through the three
  points for each physical volume. (Right) Use the results from the continuum
  extrapolation to extrapolate to infinite volume. The upper points use the
  quadratic fit to all three lattice spacing shown in the left plot, while the
  lower points use a linear fit to the two leftmost points in the left plot.
  For the upper points, we obtain , we obtain $(0.3679 \pm 0.0042) - (1.86 \pm
  0.11) / (m_{{\mu}} L)^2$. For the lower points, we obtain $(0.3608 \pm
  0.0030) - (1.92 \pm 0.08) / (m_{{\mu}} L)^2$. The errors are statistical
  only. For comparison, the analytical formula gives $0.371$.
  {\cite{Laporta:1991zw}}}
\end{figure}

The finite volume effect implies power-law corrections because of the photon
has zero mass and its propagator decreases like $1 / r^2$. To estimate the
finite volume effect in LbL, we study the amplitude as a function of spatial
momentum and time, assuming that the effect of excited states has been
controlled.

Recall the photon propagator is
\begin{eqnarray}
  G (\mathbf{k}, t_2, t_1) & = & \int \frac{\mathrm{d} p_0}{2 \pi} e^{i p_0
  (t_2 - t_1)}  \frac{1}{p_0^2 +\mathbf{k}^2} \nonumber\\
  & = & \frac{1}{2_{} | \mathbf{k} |} \exp (- | \mathbf{k} |  | t_2 - t_1 |)
  . 
\end{eqnarray}
For a single internal photon, the behavior of the integrand in small
$\mathbf{k}$ region is roughly
\begin{eqnarray}
  \int_{- \infty}^{\infty} dt_{\ensuremath{\operatorname{line}}}  \frac{1}{|
  \mathbf{k} |} \exp (- | \mathbf{k} | | t_{\ensuremath{\operatorname{loop}}}
  - t_{\ensuremath{\operatorname{line}}} |)  | \mathbf{k} | & \sim &
  \mathcal{O} \left( \frac{1}{| \mathbf{k} |} \right), 
\end{eqnarray}
where $t_{\text{line}}$ represents the time at which the photon couples to the
external line and $t_{\text{loop}}$ the location in time of the internal muon
loop, fixed by $\left( x_{\text{op}} \right)_0$. The last factor of $|
\mathbf{k} |$ comes from the fact that the photon has to couple to a neutral
loop and the coupling at such a small momentum photon is suppressed by a
factor of $| \mathbf{k} |$. Thus, the finite correction should be proportion
to
\begin{eqnarray}
  \int_0^{1 / L} \mathcal{O} \left( \frac{1}{| \mathbf{k} |} \right) d^3 k &
  \sim & \mathcal{O} \left( \frac{1}{L^2} \right) . 
\end{eqnarray}
This is precisely what we observed in the numerical study. Note that this
power-law error is caused by not including the contribution from the region
with a large separation between the muon line and the fermion loop correctly.
If we simply perform the sum over $x'$, $y'$, and $z'$ in Eq
(\ref{lbl-amp})(\ref{lbl-amp-c}) in a larger volume and reuse the point source
propagators and contractions for the fermion loop, then we would obtain a
smaller finite volume error.

\subsection{$333 \mathrm{\ensuremath{\operatorname{MeV}}}$ Pion $24^3 \times
64$ Lattice}

The computation was performed on $18$ configurations each separated by $200$
MD time unit. {\cite{Arthur:2012yc}} The muon mass is set to be $175
\mathrm{\ensuremath{\operatorname{MeV}}}$. We compute the short distance part
up to $r_{\text{max}} = 4$ in lattice unit, and sample the long distance part
with the following distribution
\begin{eqnarray}
  p_{\text{24IL}} (r) & \propto & \frac{1}{| r |^4} \exp (- 0.1 | r |) . 
\end{eqnarray}
For each configuration, $118$ pairs are used to compute the short distance
part, $128$ pairs are sampled to compute the long distance part.

Our result evaluated with muon source and sink separation $t_{\text{sep}} =
32$ is
\begin{eqnarray}
  F_2 & = & (0.0804 \pm 0.0015) \left( \frac{\alpha}{\pi} \right)^3 . 
  \label{f2-24IL-mm-0.1}
\end{eqnarray}
Because we have precise control of the distance between points $x$ and $y$, we
can plot the contribution from each point pair and bin the pairs according to
the distance $r$.

\begin{figure}[H]
  \begin{center}
    \resizebox{0.4\columnwidth}{!}{\includegraphics{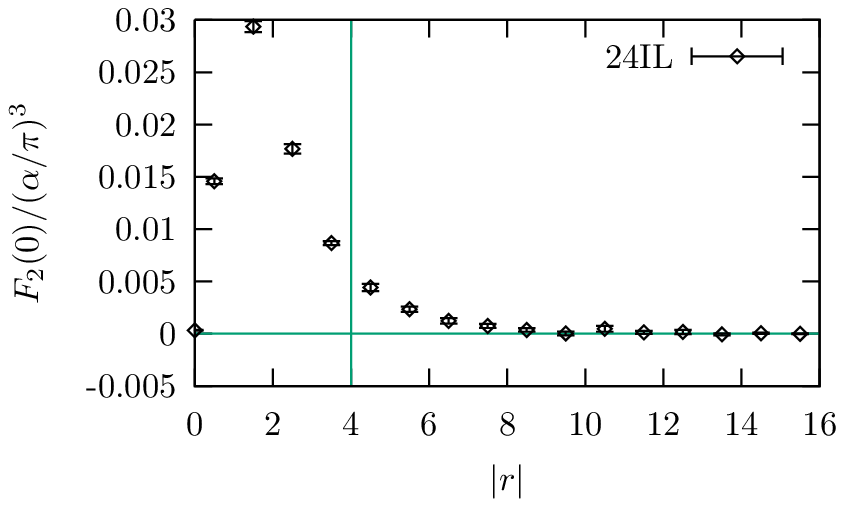}}
    \resizebox{0.4\columnwidth}{!}{\includegraphics{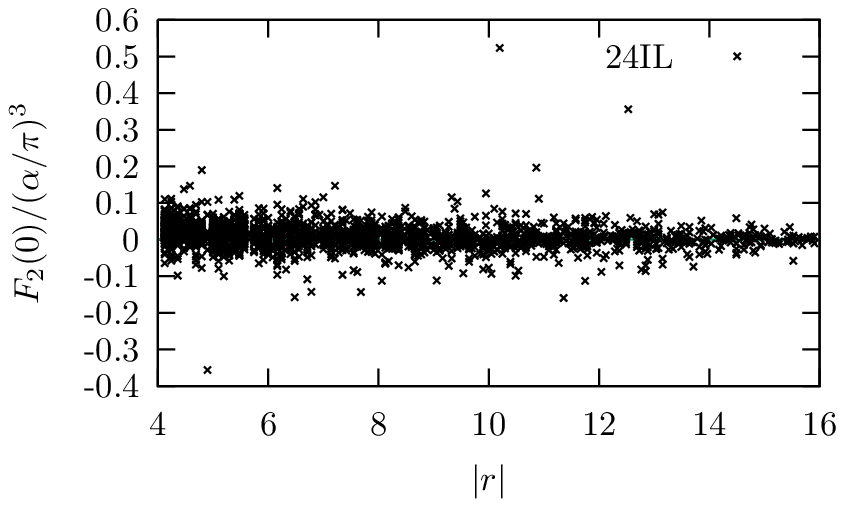}}
  \end{center}
  \caption{Results from the $24^3 \times 64$ lattice with $a^{- 1} = 1.747
  \mathrm{\ensuremath{\operatorname{GeV}}}$, $m_{\pi} = 333
  \mathrm{\ensuremath{\operatorname{MeV}}}$, $m_{{\mu}} = 175
  \mathrm{\ensuremath{\operatorname{MeV}}}$. $t_{\text{sep}} = 32$. (Left)
  Histogram of the contribution to $F_2$ from different separations $r = |x -
  y|$. The sum of all these points gives the final result for $F_2$. (Right)
  Scatter plot of results for $F_2$ for all random point pairs, adjusted by
  their sampling weight. The average value of $F_2$ from all the points gives
  the $r \ge r_{\max}$ portion of the final result. The vertical line in the
  left plot and the left-hand boundary of the points shown in the right plot
  indicate the value of $r_{\max}$. }
\end{figure}

We compare this value with results obtained in our previous attempt using the
subtraction method in Ref {\cite{Blum:2014oka}}. Not only the statistical
error becomes much smaller with the new method, the computational cost in
terms of the number of quark propagators computed, is also reduced.

\begin{figure}[H]
  \begin{center}
    \resizebox{0.5\columnwidth}{!}{\includegraphics{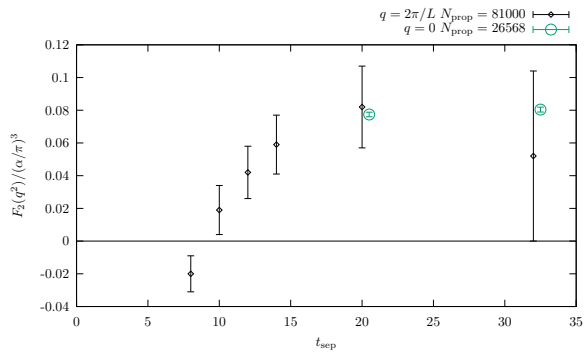}}
  \end{center}
  \caption{Results from a $24^3 \times 64$ lattice with $a^{- 1} = 1.747
  \mathrm{\ensuremath{\operatorname{GeV}}}$, $m_{\pi} = 333
  \mathrm{\ensuremath{\operatorname{MeV}}}$ using $m_{{\mu}} = 175
  \mathrm{\ensuremath{\operatorname{MeV}}}$. Results from our new methods are
  shown as red circles. The total cost is $N_{\text{prop}} = 26568$ light
  quark propagators. The small diamonds show the results from our previous
  calculation in Ref {\cite{Blum:2014oka}} computed with $N_{\text{prop}} =
  81000$ light quark propagators.}
\end{figure}

\subsection{$171 \mathrm{\ensuremath{\operatorname{MeV}}}$ Pion $32^3 \times
64$ Lattice}

With the more efficient method, we have also attempted a computation on a more
physical ensemble, performed on $23$ configurations each separated by $80$ MD
time units. {\cite{Arthur:2012yc}} The muon mass is set to be $134
\mathrm{\ensuremath{\operatorname{MeV}}}$. We compute the short distance part
up to $r_{\text{max}} = 5$ in lattice units, and sample the long distance part
with the following distribution
\begin{eqnarray}
  p_{\text{32ID}} (r) & \propto & \frac{1}{| r |^4} \exp (- 0.01 | r |) . 
\end{eqnarray}
For each configuration, $217$ pairs are used to compute the short distance
part, $512$ pairs are sampled to compute the long distance part.

\begin{figure}[H]
  \begin{center}
    \resizebox{0.4\columnwidth}{!}{\includegraphics{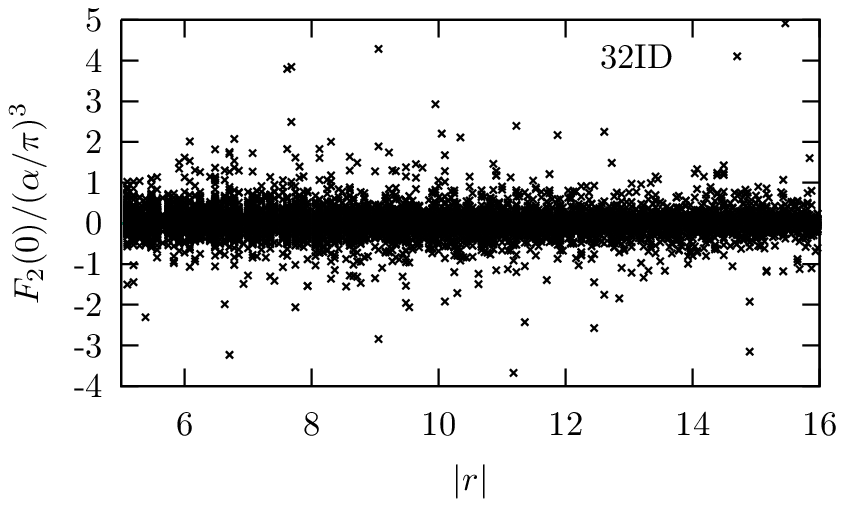}}
    \resizebox{0.4\columnwidth}{!}{\includegraphics{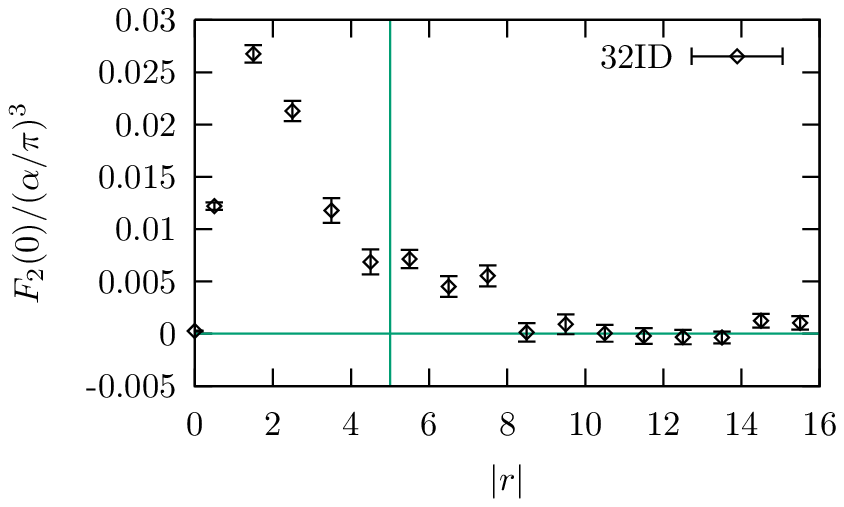}}
  \end{center}
  \caption{\label{fig-32ID-results}Results from the $32^3 \times 64$ lattice
  with $a^{- 1} = 1.371\ensuremath{\operatorname{GeV}}$, $m_{\pi} =
  171\ensuremath{\operatorname{MeV}}$, $m_{{\mu}} =
  134\ensuremath{\operatorname{MeV}}$. $t_{\text{sep}} = 32$. (Left) Scatter
  plot of results for $F_2$ for all random point pairs, adjusted by their
  sampling weight. The average value of $F_2$ from all the points gives the $r
  \ge r_{\max}$ portion of the final result. (Right) Histogram of the
  contribution to $F_2$ from different separations $r = |x - y|$. The sum of
  all these points gives the final result for $F_2$. The vertical line in the
  right plot and the left-hand boundary of the points shown in the left plot
  indicate the value of $r_{\max}$.}
\end{figure}

We use AMA technique to speed up the computations. Figure
\ref{fig-32ID-results} shows the result from the sloppy solves with $100$
iterations and $550$ low modes. The small correction term $(0.0060 \pm 0.0042)
(\alpha / \pi)^3$, which is computed separately, is then added to obtain our
final result for this lattice
\begin{eqnarray}
  F_2 & = & (0.1054 \pm 0.0054) \left( \frac{\alpha}{\pi} \right)^3 . 
  \label{f2-32ID}
\end{eqnarray}
This entire computation required $13.2$ BG/Q rack days, where one BG/Q rack is
composed of $1024$ nodes each composed of $16$ cores.

One may compare this value with the model calculation {\cite{Prades:2009tw}},
which gives $(0.08 \pm 0.02) (\alpha / \pi)^3$, although it should be noted
that Eq. (\ref{f2-24IL-mm-0.1}) and Eq. (\ref{f2-32ID}) are computed on
lattice with unphysical pion and muon mass, in finite volume, non-zero lattice
spacing, and all disconnected diagrams have been omitted.

\section{Conclusions}

We have made significant improvements to the evaluation strategy of the
connected hadronic light-by-light \ (cHLbL) diagram. With exact photon
propagators and the moment method, one can now compute the connected hadronic
light-by-light contribution (cHLbL) to $g - 2$ for the muon in the zero
momentum transfer limit directly and accurately. The muon leptonic numerical
experiments demonstrate the effectiveness of this method, and verify the
finite volume error to be \ $\mathcal{O} (1 / L^2)$. Using the improved
method, we compute the cHLbL on $24^3$ $2.71
\mathrm{\ensuremath{\operatorname{fm}}}$ lattice with $333
\mathrm{\ensuremath{\operatorname{MeV}}}$ pion and $175
\mathrm{\ensuremath{\operatorname{MeV}}}$ muon to a greater precision than in
our previous result. We also tested this method on a more close-to-physical
$32^3$ $4.6 \mathrm{\ensuremath{\operatorname{fm}}}$ lattice with a $171
\mathrm{\ensuremath{\operatorname{MeV}}}$ pion and a $134
\mathrm{\ensuremath{\operatorname{MeV}}}$ muon. We are now actively using this
method at a physical pion mass and $48^3$ $5.5
\mathrm{\ensuremath{\operatorname{fm}}}$ lattice. We also plan to address the
finite volume effect and disconnected diagrams within the framework of this
newly developed evaluation strategy.